\documentclass[10pt,A4paper,conference]{IEEEtran}

\usepackage{
            theorem, 
            amssymb, amsmath, graphicx, afterpage}

\newcommand{\RR}{\mathbb R}

\DeclareSymbolFont{BoldMath}{OML}{cmm}{b}{it}

\DeclareMathSymbol{\boa}{\mathalpha}{BoldMath}{'141}
\DeclareMathSymbol{\bob}{\mathalpha}{BoldMath}{'142}
\DeclareMathSymbol{\boc}{\mathalpha}{BoldMath}{'143}
\DeclareMathSymbol{\bod}{\mathalpha}{BoldMath}{'144}
\DeclareMathSymbol{\boe}{\mathalpha}{BoldMath}{'145}
\DeclareMathSymbol{\bog}{\mathalpha}{BoldMath}{'147}
\DeclareMathSymbol{\boh}{\mathalpha}{BoldMath}{'150}
\DeclareMathSymbol{\bom}{\mathalpha}{BoldMath}{'155}
\DeclareMathSymbol{\bon}{\mathalpha}{BoldMath}{'156}
\DeclareMathSymbol{\bop}{\mathalpha}{BoldMath}{'160}
\DeclareMathSymbol{\boq}{\mathalpha}{BoldMath}{'161}
\DeclareMathSymbol{\bor}{\mathalpha}{BoldMath}{'162}
\DeclareMathSymbol{\bos}{\mathalpha}{BoldMath}{'163}
\DeclareMathSymbol{\bot}{\mathalpha}{BoldMath}{'164}
\DeclareMathSymbol{\bou}{\mathalpha}{BoldMath}{'165}
\DeclareMathSymbol{\bov}{\mathalpha}{BoldMath}{'166}
\DeclareMathSymbol{\bow}{\mathalpha}{BoldMath}{'167}
\DeclareMathSymbol{\boxx}{\mathalpha}{BoldMath}{'170}
\DeclareMathSymbol{\boy}{\mathalpha}{BoldMath}{'171}
\DeclareMathSymbol{\boz}{\mathalpha}{BoldMath}{'172}

\DeclareMathSymbol{\boA}{\mathalpha}{BoldMath}{'101}
\DeclareMathSymbol{\boB}{\mathalpha}{BoldMath}{'102}
\DeclareMathSymbol{\boC}{\mathalpha}{BoldMath}{'103}
\DeclareMathSymbol{\boD}{\mathalpha}{BoldMath}{'104}
\DeclareMathSymbol{\boE}{\mathalpha}{BoldMath}{'105}
\DeclareMathSymbol{\boF}{\mathalpha}{BoldMath}{'106}
\DeclareMathSymbol{\boG}{\mathalpha}{BoldMath}{'107}
\DeclareMathSymbol{\boH}{\mathalpha}{BoldMath}{'110}
\DeclareMathSymbol{\boI}{\mathalpha}{BoldMath}{'111}
\DeclareMathSymbol{\boK}{\mathalpha}{BoldMath}{'113}
\DeclareMathSymbol{\boM}{\mathalpha}{BoldMath}{'115}
\DeclareMathSymbol{\boO}{\mathalpha}{BoldMath}{'117}
\DeclareMathSymbol{\boP}{\mathalpha}{BoldMath}{'120}
\DeclareMathSymbol{\boQ}{\mathalpha}{BoldMath}{'121}
\DeclareMathSymbol{\boR}{\mathalpha}{BoldMath}{'122}
\DeclareMathSymbol{\boS}{\mathalpha}{BoldMath}{'123}
\DeclareMathSymbol{\boT}{\mathalpha}{BoldMath}{'124}
\DeclareMathSymbol{\boU}{\mathalpha}{BoldMath}{'125}
\DeclareMathSymbol{\boV}{\mathalpha}{BoldMath}{'126}
\DeclareMathSymbol{\boW}{\mathalpha}{BoldMath}{'127}
\DeclareMathSymbol{\boX}{\mathalpha}{BoldMath}{'130}
\DeclareMathSymbol{\boY}{\mathalpha}{BoldMath}{'131}
\DeclareMathSymbol{\boZ}{\mathalpha}{BoldMath}{'132}

\newcommand{\bozero}{{\bf 0}}

\DeclareMathSymbol{\boalpha}{\mathalpha}{BoldMath}{'013}
\DeclareMathSymbol{\bobeta}{\mathalpha}{BoldMath}{'014}
\DeclareMathSymbol{\bodelta}{\mathalpha}{BoldMath}{'016}
\DeclareMathSymbol{\boepsilon}{\mathalpha}{BoldMath}{'042}
\DeclareMathSymbol{\bogamma}{\mathalpha}{BoldMath}{'015}
\DeclareMathSymbol{\bolambda}{\mathalpha}{BoldMath}{'025}
\DeclareMathSymbol{\bomu}{\mathalpha}{BoldMath}{'026}
\DeclareMathSymbol{\bopi}{\mathalpha}{BoldMath}{'031}
\DeclareMathSymbol{\bophi}{\mathalpha}{BoldMath}{'047}
\DeclareMathSymbol{\bosigma}{\mathalpha}{BoldMath}{'033}
\DeclareMathSymbol{\botau}{\mathalpha}{BoldMath}{'034}
\DeclareMathSymbol{\boxi}{\mathalpha}{BoldMath}{'030}
\DeclareMathSymbol{\boGamma}{\mathalpha}{BoldMath}{'000}
\DeclareMathSymbol{\boDelta}{\mathalpha}{BoldMath}{'001}
\DeclareMathSymbol{\boLambda}{\mathalpha}{BoldMath}{'003}
\DeclareMathSymbol{\boPi}{\mathalpha}{BoldMath}{'005}
\DeclareMathSymbol{\boSigma}{\mathalpha}{BoldMath}{'006}
\DeclareMathSymbol{\boPhi}{\mathalpha}{BoldMath}{'010}   
\DeclareMathSymbol{\boPsi}{\mathalpha}{BoldMath}{'011}

\newcommand{\SIR}{{\rm{SIR}}}

\newtheorem{thm}{Theorem}

\newtheorem{defi}[thm]{Definition}
\newtheorem{prop}[thm]{Proposition}

\newcommand{\argmax}{\operatorname{argmax}}

\newcommand{\calP}{{\cal P}}
\newcommand{\calPmax}{{\calP}_{\rm max}}
\newcommand{\calPSIR}{{\calP}_{\rm SIR}}
\newcommand{\diag}{{\rm diag}}

\begin{document}

\title{Optimal Power Control for Multiuser\\ CDMA Channels}

\author{\authorblockN{Anke Feiten}
\authorblockA{Institute of Theoretical Information Technology\\
RWTH Aachen University\\
52056 Aachen, Germany \\
Email: feiten@ti.rwth-aachen.de}
\and
\authorblockN{Rudolf Mathar}
\authorblockA{Institute of Theoretical Information Technology\\
RWTH Aachen University\\
52056 Aachen, Germany \\
Email: mathar@ti.rwth-aachen.de}
}

\maketitle

\begin{abstract}
In this paper, we define the power region as the set of power
allocations for $K$ users such that everybody meets a minimum
signal-to-interference ratio (SIR). The SIR is modeled in a multiuser
CDMA system with fixed linear receiver and signature sequences.
We show that the power region is convex in linear and logarithmic
scale. It furthermore has a componentwise minimal element.
Power constraints are included by the intersection with the
set of all viable power adjustments.
In this framework, we aim at minimizing the total expended power
by minimizing a componentwise monotone functional.
If the feasible power region is nonempty, the minimum is
attained. Otherwise, as a solution to balance conflicting interests, 
we suggest the projection of the minimum point in the 
power region onto the set of viable power settings.
Finally, with an appropriate utility function, the problem
of minimizing the total expended power can be seen as finding
the Nash bargaining solution, which sheds light on power assignment 
from a game theoretic point of view.
Convexity and componentwise monotonicity are essential prerequisites
for this result.
\end{abstract}

\section{Introduction}
In interference limited wireless communication systems, like
code division multiple access (CDMA), mobile users regulate
transmission power to adapt to varying radio channel and propagation
conditions.
The main purpose is to minimize interference to other users
while maintaining one´s own data rate at the lowest possible energy
consumption. A number of recent papers is dealing with the
intertwining effect of power adjustment and feasible data rates per
user in wireless networks, hence defining a concept of
overall network capacity, 
see e.g., \cite{BoSt05, ImMa03, ImMa05, TseHan99, ViAnTse99}, 
and references therein.

A companion problem is the design of efficient power control algorithms,
preferably such that each user needs only local information 
to update his power settings, but
global convergence is assured, see \cite{CaImMa04, Han95, MeHe01, Ya95}.

Refined models also include stochastic fading effects of the channel.
The usual approach here is to minimize power consumption subject
to certain outage probability constraints. Appropriate models,
structural properties of the corresponding feasible region, adequate
power control algorithms and their convergence are investigated
in \cite{CaImMa04, KaBo02, PaEvDe03, PaEvDe05, UlYa98, VaDa02}.

In this paper, we investigate properties of the power region
and the existence of energy minimal power settings assuming known channel
information and fixed signature and linear receiver sequences
in CDMA systems. To summarize, the main contributions are as follows.

In Section \ref{systemmodel}
we first define the power region for $K$ users as the set of power 
settings $\bozero\le\bop\in\RR^K$, such that in a community of
$K$ users each user $i$ encounters a signal-to-noise ratio 
above a certain threshold $\gamma_i$.
The power region is shown to be convex in linear and logarithmic scale.
It furthermore contains a componentwise minimal element $\bop^*$ that can be
explicitly determined. 

In Section \ref{energyefficient} energy efficient power allocation is
formalized as minimizing a componentwise monotone function $h$
over the power region.  
In practice, however, power is limited. This is included in our 
model by introducing the feasible power region as the intersection with
a convex and downward closed set of viable power adjustments.
In the case that there is no feasible power allocation, we suggest
the projection of $\bop^*$ onto the viable power adjustments as 
a solution which balances between conflicting interests of users.
The solution can be computed by applying cyclic projections 
onto simple affine subsets.

Finally, by use of an appropriate utility function we show in Section 
\ref{gamesection} how optimal power allocation can be 
interpreted as a cooperative game.
It turns out that the Nash bargaining solution coincides with the
solution of the original power minimization problem.

\section{System Model} 
\label{systemmodel}
In a synchronous multiuser CDMA communication system with $K$ users
and processing gain $N$ let $\bos_i\in\RR^N$, $i=1,\ldots,K$, denote the 
$N$-dimensional signature sequence of user $i$. Let
$G_{ij}$ denote the fixed path gain from user $j$ to the assigned base 
station of user $i$. Usually $G_{ij}$ is subject to slow fading effects
which are assumed to be known to the transmitter.
Suppose the symbol of user $i$ is decoded using a linear receiver
represented by some vector $\boc_i\in\RR^N$. The signal-to-interference
ratio of user $i$ is then given as
\[
\SIR_i(\bop)=\frac{G_{ii}(\boc_i'\bos_i)^2 p_i}
            {\sum_{j\not=i}G_{ij}(\boc_i'\bos_j)^2 p_j+\sigma^2 (c_i' c_i)^2},
\]
where $\sigma^2$ denotes the variance of the additive Gaussian noise
and $\bop=(p_1,\ldots,p_K)$ the vector of transmit powers.
In the following we assume that 
the receiver sequences $\boc_i$ are fixed. Summarizing the known channel 
and receiver effects into $A_{ij}=G_{ij}(\boc_i'\bos_j)^2$
we obtain the following $\SIR_i$ of user $i$
\[
\SIR_i(\bop)=\frac{A_{ii} p_i}
             {\sum_{j\not=i}A_{ij} p_j + C_{ii}\sigma^2}.
\]
with $C_{ii}=(\boc_i'\boc_i)^2$.
Now given quality-of-service requirements $\gamma_1,\ldots,\gamma_K$ for
each user, we define the {\em power region} 
$\calPSIR$ as the set of power settings $\bop\in\RR^K$ such that 
each user $i$ meets his minimum SIR requirement $\gamma_i$, i.e.,
\begin{equation}\label{powerregion}
{\calPSIR}=\big\{\bop\ge \bozero\mid \SIR_i(\bop)\ge\gamma_i,\ 
               i=1,\ldots,K\big\}.
\end{equation}
Here and in the following orderings '$<$' and '$\le$' between vectors
are always meant componentwise.
Obviously it may happen that not all requirements $\gamma_i$ can be 
simultaneously satisfied in which case $\calPSIR$ is empty.

We now prove convexity and log-convexity of 
the power region $\calPSIR$. The set $\calPSIR$ is called log-convex,
if for any $\bop^{(1)},\,\bop^{(2)}\in\calPSIR$ and any $0\le\alpha\le 1$
the point
\[
\bop^{(\alpha)}={\bop^{(1)}}^\alpha {\bop^{(2)}}^{1-\alpha}\in{\calPSIR},
\]
where powers $\bop^\alpha=(p_1^\alpha,\ldots,p_K^\alpha)$
are applied componentwise. Taking logarithms componentwise gives
\[
\log\bop^{(\alpha)}=\alpha\log{\bop^{(1)}}
 +(1-\alpha)\log{\bop^{(2)}}
\]
which means that the set $\calPSIR$ is convex in logarithmic scale.

\begin{prop} \label{powerregconv}
The power region $\calPSIR$ is convex and log-convex.
\end{prop}

\begin{proof}
Consider the sets
\[
{\cal P}_i=\big\{\bop\mid A_{ii}p_i-\gamma_i\sum_{j\not=i} 
A_{ij} p_j\ge \gamma_i C_{ii}\sigma^2\big\},\ i=1,\ldots,K,
\]
which are closed convex affine halfspaces in $\RR^K$.
Obviously, $\calPSIR=\bigcap_{i=1}^K {\cal P}_i$, and from
Theorem C in section III of \cite{RoVa73} it follows that
$\calPSIR$ is a closed and convex polytope.

To prove log-convexity we show that
\begin{equation} \label{logconveq1}
\begin{split}
\SIR_i(\bop^{(\alpha)})
  &\ge \big( \SIR_i(\bop^{(1)})\big)^\alpha 
   \big( \SIR_i(\bop^{(2)})\big)^{1-\alpha}\\
  &\ge\gamma_i^\alpha \gamma_i^{1-\alpha}=\gamma_i
\end{split}
\end{equation}
for all $i=1,\ldots,K$, which entails $\bop^{(\alpha)}\in\calPSIR$.
The first inequality in (\ref{logconveq1}) follows from
H\"older's inequality since
\begin{align*}
&\sum_{j\not=i}A_{ij}{p_j^{(1)}}^\alpha {p_j^{(2)}}^{1-\alpha}
+C_{ii}\sigma^2\\
&\quad\le \Big(\sum_{j\not=i}A_{ij} p_j^{(1)} + C_{ii}\sigma^2\Big)^\alpha
     \Big(\sum_{j\not=i}A_{ij} p_j^{(2)} + C_{ii}\sigma^2\Big)^{1-\alpha},
\end{align*}
hence yielding
\begin{align*}
&\SIR_i\big(\bop^{(\alpha)}\big)\\
&\ =\frac{A_{ii}{p_i^{(1)}}^\alpha{p_i^{(2)}}^{1-\alpha}}
       {\sum_{j\not=i}A_{ij}{p_i^{(1)}}^\alpha{p_i^{(2)}}^{1-\alpha}
        +C_{ii}\sigma^2}\\
&\ \ge\Big(\frac{A_{ii}{p_i^{(1)}}}
       {\sum_{j\not=i}A_{ij}{p_j^{(1)}}+C_{ii}\sigma^2}\Big)^\alpha
    \Big(\frac{A_{ii}{p_i^{(2)}}}
       {\sum_{j\not=i}A_{ij}{p_j^{(2)}}+C_{ii}\sigma^2}\Big)^{1-\alpha}\\
&\ =\big( \SIR_i(\bop^{(1)})\big)^\alpha 
   \big( \SIR_i(\bop^{(2)})\big)^{1-\alpha}
\end{align*}
for all $i=1,\ldots,K$.
\end{proof}

\section{Energy Efficient Power Allocation}
\label{energyefficient}
For convenience of notation we quote the following result 
from \cite{ImMa05}. It deals with solutions of the equation
\begin{equation}\label{iaxc}
[\boI-\boA]\boxx=\boc
\end{equation}
when $\boA$ is a non-negative but not necessarily irreducible matrix. 
The proof given in \cite{ImMa05} is direct and self-contained, and 
does not rely on the Perron-Frobenius theory. Let 
$\rho(\boA)$ denote the spectral radius of a square matrix $\boA$.

\begin{prop} \label{starterlemma}
Let $\boA\in\RR^{n\times n}$ be non-negative.
\begin{itemize}
\item[{\rm a)}] If there are $\boxx>\bozero$\,, $\boc>\bozero$ satisfying
{\rm (\ref{iaxc})}, then $\rho(\boA)<1$. 
\item[{\rm b)}] If $\rho(\boA)<1$, then $\boI-\boA$ is non-singular and 
for every $\boc>\bozero$, the
unique solution $\boxx\in\RR^n$ of {\rm (\ref{iaxc})} is positive. 
\item[{\rm c)}] If $\rho(\boA)<1$, then for every $\boc\geq\bozero$, the 
unique solution $\boxx\in\RR^n$ of {\rm (\ref{iaxc})} is non-negative.
\item[{\rm d)}] If $\boc>\bozero$ and there exists $\boy>\bozero$
such that $[\boI-\boA]\boy\geq\boc$, then {\rm (\ref{iaxc})} has a unique solution 
$\boxx$ and $\bozero<\boxx\leq\boy$.
\end{itemize}
\end{prop}

The above is now applied to $\calPSIR$. The inequalities defining
(\ref{powerregion}) can be rewritten 
as a system of linear inequalities. For this purpose write
$\boB=(b_{ij})_{i,j=1}^K$, with
\[
b_{ij}=\begin{cases} {A_{ij}}/{A_{ii}}, & i\neq j,\\
0, & i=j,
\end{cases}
\]
and $\botau=(\tau_1,\dots,\tau_K)'$, where $\tau_i=C_{ii}\sigma^2/A_{ii}$.
Then for every $\bop>\bozero$ it holds that $\bop\in\calPSIR$
if and only if
\begin{equation}\label{linineq}
\left[\boI-\diag(\bogamma)\boB\right]\bop\geq \diag(\bogamma)\botau,
\end{equation}
where $\diag(\bogamma)$ denotes the matrix with diagonal entries 
$\gamma_i$ and nondiagonal entries equal to zero.

If system (\ref{linineq}) has a solution $\bop>0$, then
there is a unique solution $\bop^*\le\bop$ satisfying
\begin{equation}\label{lineq}
\left[\boI-\diag(\bogamma)\boB\right]\bop^*= \diag(\bogamma)\botau,
\end{equation}
as follows from Proposition \ref{starterlemma}. 
Moreover, for any given $\bogamma>\bozero$, the equation 
$\left[\boI-\diag(\bogamma)\boB\right]\bop=\diag(\bogamma)\botau$
has a positive solution $\bop$ if and only if the spectral radius
$\rho(\diag(\bogamma)\boB)<1$,
and in that case, the solution is unique.
Denote it by $\bopi(\bogamma)=
(\pi_1(\bogamma),\dots,\pi_K(\bogamma))'$. Thus
\begin{equation}\label{piofgamma}
\bopi(\bogamma)=\left[\boI-\diag(\bogamma)\boB\right]^{-1}
\diag(\bogamma)\botau
\end{equation}
with all components positive.

Summarizing our results so far, we have the following

\begin{prop} \label{uniformmin}
If $\calPSIR\not=\emptyset$,
then there is a unique power allocation $\bop^*=\bopi(\bogamma)$ such that
\begin{align*}
\SIR_i(\bop^*)&=\gamma_i \quad\text{ for all } i=1,\ldots,K \text{ and }\\
\bop^*&\le\bop \quad\text{ for all } \bop\in\calPSIR.
\end{align*}
\end{prop}

Energy efficient power allocation can be formalized with the help
of some function $h:\RR^K_+\to\RR$ as 
the following optimization problem.

\begin{equation}
\text{minimize } h(\bop) \text{ over all } \bop\in\calPSIR
\end{equation}

From Proposition \ref{uniformmin} it is clear that for any
componentwise monotone function $h$ the minimum is attained at
$\pi(\bogamma)$ whenever $\calPSIR\not=\emptyset$.
Examples of such functions $h$ are 
\[
h(\bop)=\Vert\bop\Vert_q=
\Big(\sum_{i=1}^K |p_i|^q\Big)^{1/q},
\]
the $\ell_q$-norms, $q\ge1$,
with the special case $h(\bop)=\sum_{i=1}^K p_i$ for $q=1$.

In practice, however, power is limited. Hence, mobiles may select their power
adjustment only from a bounded set $\calPmax$, say. In the following we
assume that $\calPmax$ is convex and closed under 
simultaneous decrease of power (see \cite{ImMa05}), i.e.,
\begin{equation}\label{poweringdown}
\text{if } \bop\in\calPmax \text{ and }\bozero<\boq\le\bop,
\text{ then } \boq\in\calPmax.
\end{equation}
Typical examples of structure (\ref{poweringdown})
are individual power constraints
\begin{equation}\label{indpow}
0 \le p_i \le p_{i, {\rm max}}, \quad i=1,\ldots,K,
\end{equation}
or a limited total power budget as
\begin{equation}\label{totalpow}
\sum_{i=1}^K p_i\le p_{\rm max},\quad
p_i \ge 0,\ i=1,\ldots,K,
\end{equation}
or a combination hereof by intersecting both sets.

Under constraints (\ref{indpow}) a relevant example of a componentwise 
monotone function is
\begin{equation} \label{gamethfct}
h(\bop)=
-\prod_{i=1}^K\big(e^{p_{i,{\rm max}}}- e^{p_i}\big)
\end{equation}
Its meaning will be clear from a game theoretic interpretation of power
allocation with certain utility functions in Section \ref{gamesection}.

Energy efficient {\em feasible} power allocation can now be written as
\[
\text{minimize } h(\bop)\text{ over all }
\bop\in\calPSIR\cap\calPmax,
\]
where by assumption $\calPSIR\cap\calPmax$ is a convex subset of $\RR^K$.

If $\calPSIR\cap\calPmax\not=\emptyset$ it follows from
Proposition \ref{starterlemma} and (\ref{poweringdown})
that $\bopi(\bogamma)$ from (\ref{piofgamma})
is the optimal power allocation for any componentwise 
monotone function $h$.

In the important case that $\calPSIR\not=\emptyset$ but
$\calPSIR\cap\calPmax=\emptyset$ there exists no feasible
power allocation to satisfy all SIR requirements 
simultaneously. A solution $\hat\bop\in\calPmax$
which balances the conflicting interests of users is the projection of 
$\pi(\bogamma)$ onto the convex set $\calPmax$,
\[
\hat\bop={\rm Proj}\big(\bopi(\bogamma)\mid \calPmax\big).
\]
As follows from the classical projection theorem for Hilbert 
spaces (see \cite{Lue69}), $\hat\bop$ is unique. It
represents the feasible power adjustment coming closest
to the required but infeasible $\bop^*=\bopi(\bogamma)$.

For the above constraints (\ref{indpow}) and (\ref{totalpow})
$\hat\bop$ can be computed by a convergent cyclic projection algorithm 
as investigated in \cite{GaMa89}. 
In this approach, starting with $\bop^*$, points are iteratively 
projected onto convex sets whose intersection forms the set where the 
projection onto is sought. Here only projections onto affine halfspaces
of the form ${\cal H}=\{\bop\mid\boa'\bop\le\beta\}$
are needed. The general solution in this case is given by
\[
{\rm Proj}(\bop\mid {\cal H})
=\bop-\frac{1}{\boa'\boa}\big(\boa'\bop-\beta\big)^+\boa.
\]
By selecting $\boa=\boe_i$ (the $i$-th unit vector) and
$\beta=p_{i,{\rm max}}$ we obtain the projection onto
the set (\ref{indpow}). The choice 
$\boa=(1,\ldots,1)'$ and $\beta=p_{\rm max}$ yields the projection onto
(\ref{totalpow}).

We proceed by interpreting optimal power allocation as
a cooperative game, and by embedding this approach into the
above framework.

\section{Power Allocation as a cooperative game}
\label{gamesection}
\begin{figure}[t]
\begin{center}
\begin{minipage}[t]{6cm}
\includegraphics[width=0.9\linewidth]{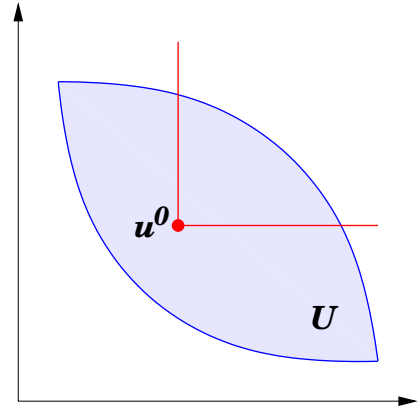} 
\caption{\label{figNBS}A Nash bargaining problem}
\end{minipage} 
\end{center}
\end{figure}

We start by reviewing some basic concepts of cooperative bargaining 
theory \cite{Mu99}.
A $K$-person bargaining problem is a pair $(U,\bou^0)$, 
where $U \subset \RR^K$ is a 
nonempty convex, closed and upper bounded set and 
$\bou^0=(u^0_1, \dots, u^0_K)\in \RR^K$ such that 
$\bou\geq \bou^0$ componentwise for some $\bou=(u_1, \dots, u_K) \in U,$ 
see Fig.~\ref{figNBS}. 

The elements of $U$ are called outcomes and $\bou^0$
is the disagreement outcome.
The interpretation of such a problem is as follows. A community of
$K$ bargainers is faced with the problem to negotiate for a fair point
in the convex set $U.$ If no agreement can be achieved by the bargainers, 
the disagreement utilities $u^0_1,\dots,u^0_K$ will be the outcome of the game.
Let $B_K$ denote the family
of all $K$-person bargaining problems. \\
A bargaining solution is a function $F: B_K \to \RR^K$ such that $F(U,\bou^0)\in U$
for all $(U,\bou^0)\in B_K.$ 
Nash suggested a solution that is based on four axioms, as given below.
\vspace{5pt}

\noindent
(WPO) \textit{Weak Pareto optimality}:\\ 
$F: B_K \to \RR^K$ is called weakly Pareto
optimal, if for all $(U,\bou^0)\in B_K$ there exists no 
$\bou \in U$ satisfying  $\bou > F(U,\bou^0)$.
\vspace{0pt}

\noindent
(SYM) \textit{Symmetry}:\\ 
$F: B_K \to \RR^K$ is symmetric if 
$F_i(U,\bou^0)=F_j(U,\bou^0)$
for all $(U,\bou^0)\in B_K$ that are symmetric with respect to a subset 
$J\subseteq \{1,\dots ,K\}$ for all $i,j \in J$ (i.e., $u_i^0=u_j^0$ and  
$(u_1,u_2,\dots , u_{i-1},u_j,u_{i+1}, \dots , u_{j-1},u_i,u_{j+1}, \dots 
\dots,u_K)\in U$ for all $i<j \in J, \bou \in U$).
\vspace{5pt}

\noindent
(SCI) \textit{Scale covariance}:\\ 
$F: B_K \to \RR^K$ is scale covariant if 
$F(\varphi(U), \varphi(\bou^0))=\varphi(F(U,\bou^0))$ for all 
$\varphi: \RR^K \to \RR^K, \varphi(\bou )=\bar{\bou}$ with 
$\bar{u}_i=a_iu_i+b_i$,  $a_i,b_i \in \RR$, $a_i>0,\ i=1,\dots ,K$.
\vspace{5pt}

\noindent
(IIA) \textit{Independence of irrelevant alternatives}:\\ 
$F: B_K \to \RR^K$ is
independent of irrelevant alternatives, if $F(U,\bou^0)=F(\bar{U},\bar{\bou}^0)$ 
for all $(U,\bou^0),(\bar{U},\bar{\bou}^0)\in B_K$ with 
$\bou^0=\bar{\bou}^0,\ U \subseteq \bar{U}_0$ and $F(\bar{U},\bar{\bou}^0)\in U$.
\vspace{5pt}

\noindent
{\bf Remark. }
Weak Pareto optimality means that no bargainer can gain over the solution outcome.
Symmetry, scale covariance and independence of irrelevant alternatives are the so called axioms of fairness. The symmetry property states that the solution does not depend on the specific label, i.e., users with both the same initial points and objectives will obtain the same performance. Scale covariance requires the solutions to be covariant under positive affine transformations. Independence of irrelevant alternatives demands that the solution outcome does not change when the set of possible outcomes shrinks but still contains the original solution.\\
These four axioms imply \textit{Pareto optimality} which means that it is impossible to increase any player's
utility without decreasing another player's utility.
\vspace{5pt}

The Nash bargaining solution is defined as follows.

\begin{defi} \label{defNBS}
A function $N:B_K \to \RR^K$ is said to be a Nash bargaining solution (NBS) if
\begin{eqnarray*}
N(U,\bou^0)=\argmax\Big\{\prod_{\substack{ 1\leq j \leq K \\ u_j \neq u_j^0 }} 
\left(u_j-u^0_j\right) \left| \bou \in U, \bou \geq \bou^0 \right. \Big\},
\end{eqnarray*}
whenever $U \backslash \{\bou^0\} \neq \varnothing$ and $N(U,\bou^0)=\bou^0,$ 
otherwise. 
\end{defi}

\begin{figure}
\begin{center}
\begin{minipage}[t]{6cm}
\includegraphics[width=0.9\linewidth]{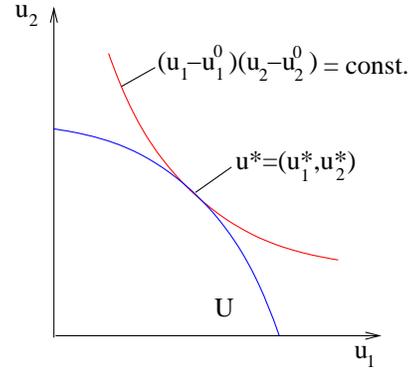} 
\caption{\label{sets3}The Nash Bargaining Solution}
\end{minipage} 
\end{center}
\end{figure}

The NBS aims at maximizing the product of the
users' gain from cooperation, see Fig.~\ref{sets3} for the two dimensional case.
In addition it is uniquely characterized by the
four axioms stated above. The proof of the following theorem can be found in \cite{StSt84}.

\begin{thm}
Let $F: B_K \to \RR^K$ be a bargaining solution. Then the following two statements are equivalent: 
\begin{enumerate}
\item[(a)] $F=N.$ 
\item[(b)] $F$ satisfies WPO, SYM, SCI, IIA.
\end{enumerate}
\end{thm}

In the following we consider the feasible power region 
\[
{\cal P}=\calPSIR\cap\calPmax
\]
with $\calPmax=\{\bop>\bozero\mid p_i\le p_{i,{\rm max}}\}$, as defined in
\eqref{powerregion} and (\ref{indpow}). 

Our aim is to single out one element of the power region ${\cal P}.$ 
There are many solution concepts to choose a reasonable element of ${\cal P},$ 
e.g. bargaining theory, proportional fairness, max-min fairness. 
We choose bargaining theory and the NBS in accordance
with the above axioms.

Transmit power of each mobile station is bounded by $p_{i,{\rm max}}$. 
Due to limited battery power each user aims at using 
the lowest power possible.
Hence we introduce a utility function $f_i$ for each mobile station $1\leq i\leq K$
as follows
\begin{eqnarray*} 
f_i: {\cal P} \to \RR, \quad  \bop \mapsto e^{p_{i,{\rm max}}}-e^{p_i}.
\end{eqnarray*}
The task now is to find an element of the feasible power region such that 
the utility of each player is maximized. This task however is impossible to 
solve. As an alternative, we need to find an element in the utility set 
$f({\cal P})$ that is superior to other elements. The utility set is 
defined as the image of the utility functions
\begin{eqnarray*} 
f({\cal P})=\{ e^{\bop_{\rm max}}-e^{\bop} \mid \bop \in \calP  \},
\end{eqnarray*}
where $e^{\bop}=(e^{p_1},\dots,e^{p_K})$ is defined componentwise and 
$e^{\bop_{\rm max}}=(e^{p_{1,{\rm max}}},\ldots,e^{p_{K,{\rm max}}})$
accordingly.

Clearly we should choose a Pareto optimal element. The question arises at
which of the infinitely many Pareto optimal points the system should be
operated. 
From the perspective of resource sharing, one of the natural criteria is the
notion of fairness. This, in general is a loose term and there are many notions of fairness. 
One of the commonly used notions is that of max-min fairness which penalizes large users.
Max-min fairness corresponds to a Pareto optimal point. However, it is not easy to take into
account that users might have different requirements. A much more satisfactory approach
is the use of fairness from game theory as introduced above. Another common solution concept of
fairness is proportional fairness. It can be shown that proportional fairness 
leads in fact to the NBS. Therefor we confine ourselves to a game theoretic 
approach here.

\begin{figure}[t]
\begin{center}
\begin{minipage}[t]{6cm}
\includegraphics[width=0.9\linewidth]{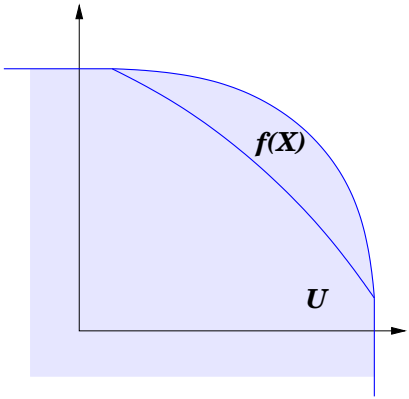} 
\caption{\label{sets2}The extension $U$ of the utility set $f(\calP)$}
\end{minipage} 
\end{center}
\end{figure}

In our cooperative game, players are formed by mobile stations, and they have 
to agree upon some element of the utility set $f(\calP).$ 
The bargaining set $U$ is now obtained by extending $f(\calP)$ to
\begin{eqnarray} \label{defU}
U=\left\{ \bou \in \RR^K \mid \exists \, \bop \in \calP  \text{ s.t.~ } \bou 
\leq  f(\bop) \right\}.
\end{eqnarray}
As is shown in \cite{FeMa04} the convexity of $U$ follows since 
$f_i$ are concave functions. Fig.~\ref{sets2} generically depicts 
the utility set and its extension.
Observe that the outcome of the cooperative game lies in $f(\calP)$, 
so that the enlargement to $U$ is mainly of technical reasons to 
assure convexity.

Moreover
\begin{eqnarray} \label{defu0}
\bou^0=-e^{\bop_{\rm max}}
\end{eqnarray}
represents the disagreement outcome, where each user has to transmit with 
its maximum power if the mobile stations fail to achieve an agreement.
In summary, $(U,\bou^0)$ is a K-person bargaining game.
In the following the NBS of this game is determined. It turns out that 
the NBS coincides with the previously defined minimum power solution
under function $h$ in \eqref{gamethfct}. The proof follows easily from
Definition~\ref{defNBS}.


\begin{prop}
The unique NBS to the bargaining problem $(U,\bou^0)$ defined in 
\eqref{defU} and \eqref{defu0} is the solution to the following 
optimization problem:
\begin{equation}\label{bargainproblem}
\max \prod_{i=1}^{K} (e^{ p_{i,{\rm max}}}-e^{p_i})
\end{equation}
{such that}
\begin{align*}
\SIR_i(\bop)&\ge\gamma_i, \\
p_i &\geq 0, \\
p_{i,{\rm max}}-p_i&\geq 0,\quad 1=1,\ldots,K. 
\end{align*}
\end{prop}
\vspace{1em}

Problem~\eqref{bargainproblem} is equivalent to minimizing 
function \eqref{gamethfct} subject to constraints \eqref{indpow}.
If the constraining set is nonempty, the solution is given
by $\bop^*=\bopi(\bogamma)$ from (\ref{piofgamma}), as is shown in 
Section \ref{energyefficient}.



\section{Conclusion}
This paper deals with the power control problem for a multiuser
CDMA channel. It is shown that the power region is both
convex and log-convex. It furthermore contains a uniformly 
minimal element $\bop^*$, at which any componentwise montone function
attains its minimum. If there is no feasible power allocation
we suggest the point of minimum distance to $\bop^\ast$ in the viable 
power region as a solution to the feasible minimum power problem.
This point can be easily computed by a cyclic projection algorithm.
The paper concludes with showing that for an appropriate 
utility function the minimum power problem with restricted power budget 
is obtained as the Nash bargaining solution for an adaptively defined
cooperative game.

\section*{Acknowledgment}
This work was supported by DFG grant Ma 1184/11-3.
\vspace{1em}




%

\bibliographystyle{IEEEtran.bst}
\bibliography{cdma}

\end{document}